\documentstyle{article}
\font\tenrm=cmr10
\font\tenit=cmti10
\font\elevenbf=cmbx10 scaled\magstep 1
\font\elevenrm=cmr10 scaled\magstep 1

\newcommand{\be}{\begin{equation}}
\newcommand{\ee}{\end{equation}}
\renewenvironment{thebibliography}[1]
 { \elevenrm
   \begin{list}{\arabic{enumi}.}
    {\usecounter{enumi}     \setlength{\parsep}{0pt}
     \setlength{\itemsep}{3pt} \settowidth{\labelwidth}{#1.}
     \sloppy
    }}
     {\end{list}}

\begin{document}
\begin{center}{\elevenbf
Approximate Analytical Solutions of the Baby Skyrme Model}\\
\vglue 0.5cm
{\tenrm T.A. Ioannidou$^1$, V.B. Kopeliovich$^2$, W.J. Zakrzewski$^3$ } \\
\vglue 0.3cm
$^1${\tenit Institute of Mathematics,  University of Kent,
Canterbury CT2 7NF, UK }\\
$^2${\tenit Institute for Nuclear Research of Russian Academy of Sciences,
Moscow 117312, Russia}\\
$^3${\tenit  Department of Mathematical Sciences, University of Durham,
Durham DH1 3LE UK }\\
\end{center}
\vglue 0.3cm
{\rightskip=2pc
 \leftskip=2pc
\tenrm\baselineskip=11pt
 \noindent
In this paper we show that many properties of the baby skyrmions,
which have been determined numerically, can be understood
in terms of an analytic approximation.
In particular, we show that the approximation captures properties of the
multiskyrmion solutions (derived numerically) such as their stability
 towards decay into various channels, and that it is more accurate for the
 ``{\it new baby Skyrme model}"  which describes anisotropic physical
systems in terms of multiskyrmion fields with axial symmetry. Some universal
characteristics of configurations of this kind are demonstrated, which do not
depend on their topological number.
\vglue 0.6cm}
\elevenrm
\baselineskip=14pt

\section{Introduction}

It is known that the two-dimensional O(3) $\sigma$-model \cite{1} possesses
metastable states which when perturbed can shrink or spread out
due to the conformal (scale) invariance of the model \cite{2,3,4}.
This implies that the metastable states can be of any size and so
a fourth order in derivatives term, the so-called Skyrme term,
needs to be added to break the scale invariance of the model.
However the resulting energy functional has no minima and
a further extra term  is needed to stabilize the size of the
corresponding  solitons, ie a term which contains no derivatives
of the field, often called the potential (or mass) term.
In this case the field can be thought of as the magnetization vector
of a two-dimensional ferromagnetic substance \cite{1},
and  the potential term describes the coupling of the magnetization vector
to a constant external magnetic field.
As the extra terms contribute to the masses of the solitons
their dependence deviates from a simple law in which the
skyrmion mass is proportional to the skyrmion (topological) number
and the two-skyrmion configuration becomes stable showing that the model
possesses bound states \cite{5}.

In this paper we demonstrate  that the simple analytical method used for
the description of the three-dimensional Skyrme model presented in \cite{6}
can be used also to study various properties of the low-energy
states of the corresponding two-dimensional $\sigma$-model when
the parameters which determine  the contributions  of the Skyrme
and the  potential term are not large.
More precisely, it was possible to describe analytically the basic
properties of the three-dimensional skyrmions for large baryon numbers
\cite{6}, and so it is worthwhile to derive such a  description
for the two-dimensional  O(3) $\sigma$-model as well.
In general, such analytical discussions of soliton models
are useful as they lead to a  better understanding of the
soliton properties.
The two-dimensional O(3) $\sigma$-model is widely used  to describe
ferromagnetic systems, high-temperature superconductivity, etc and
so the results  obtained here can be useful for the understanding
 of these phenomena.

Our method is based on the  ansatz introduced in \cite{6} and is
accurate for the so-called ``{\it new baby Skyrme model}"
\cite{7} which describes  anisotropic physical systems.
In fact, its accuracy increases as the skyrmion number $n$ increases, and this
method allows to predict some universal properties of the ring-like 
configurations for large $n$, independently on its particular value.
Although such models are not integrable, in the case where $n$ is large
the ``{\it new baby Skyrme model}"  appears to have the properties of an
integrable system.

\section{Near the Nonlinear $O(3)$ $\sigma$-Model}

The Lagrangian density of the $O(3)$ $\sigma$-model with the additional terms
introduced and discussed in \cite{5,7,8}
is:\footnote{The first few paragraphs of this section follow quite closely to
\cite{5,8} and are included to make the paper more selfconsistent. }
\be
{\cal L} = {g^2\over 2}\left(\partial_\alpha \vec{n}\right)^2
-{1\over 4e^2}\left[\partial_\alpha\vec{n},\partial_\beta\vec{n}\right]^2
-g\sp2V.
\label{Lag}
\ee
Here $\partial_\alpha=\partial/\partial x^\alpha$;
 $x^\alpha$, $\alpha=0,1,2$, refer to both time and
spatial components of $(t,x,y)$; and  the field $\vec{n}$ is a scalar field
with three components $n_a$, $a=1,2,3$, satisfying the condition
$\vec{n}^2=n_1^2+n_2^2+n_3^2=1$.
The constants $g$, $e$ are free parameters, ie $g\sp2$ has the
dimension of energy. It is useful to think of $g^2$ and $1/ge$ as 
natural units of energy and length respectively. 
The first term in (\ref{Lag}) is familiar from $\sigma$-models;
the second term, fourth order in derivatives, is the analogue of the
Skyrme term; while the last term is the potential term.
In fact, the potentials for the  ``{\it old baby Skyrme model}" (OBM)
and the  ``{\it new baby Skyrme model}" (NBM) describing anisotropic systems
are given by
\begin{eqnarray}
V_{_{\tiny\mbox{OBM}}}&=&\mu^2\left(1-n_3\right), \nonumber \\
V_{_{\tiny\mbox{NBM}}}&=&{1\over 2}\,\mu^2 \left(1-n_3\sp2\right)
\label{poten}
\end{eqnarray}
respectively, and $\mu$ has the dimension of energy, so
$1/\mu$ defines a second length scale in our model. Evidently,
$V_{_{\tiny\mbox{NBM}}} \leq V_{_{\tiny\mbox{OBM}}}$ at fixed value of $\mu$.

In three spatial dimensions the Skyrme term is necessary for
the existence of soliton solutions, but the inclusion of a
potential is optional from the mathematical point of view.
Physically, however, a potential of a certain form is required to give the
pions a mass \cite{9}.
By contrast, in two dimensions a potential term has to be included in the above
Lagrangian in order soliton solutions to exist.
As it has been shown in \cite{10}, the different potential terms give quite 
different properties to the  multiskyrmion configurations when
the skyrmion number is large. Our analytical treatment here supports this 
conclusion, as shown in sections 3-5.

We are only interested in configurations with finite energy, so we define
the configuration space to be the space of all maps $\vec{n}:
 R^2 \rightarrow S^2$ which tend to the constant field $(0,0,1)$
(so-called vacuum) at spatial infinity
\be
\lim_{|x| \rightarrow \infty} \vec{n}(\vec{x})=(0,0,1).
\ee
Thus every configuration $\vec{n}$ may be regarded as a
representative of a homotopy class in $\pi_2(S^2)=\bf{Z}$
and has a corresponding integer degree of the form
\be
\deg\left[\vec{n}\right]=\frac{1}{8\pi} \int d^2x\,
 \epsilon^{ab}\, \vec{n}\, \left(\partial_{a} \vec{n} \times \partial_{b}
\vec{n}\right).
\ee
The vacuum field is invariant under the symmetry group
$G=E_2\times SO(2)_{iso} \times P$, where $E_2$ is the
Euclidean group of translations and rotations in two dimensions
which acts on fields via pull-back.
$SO(2)_{iso}$ is the subgroup of the
three-dimensional rotation group acting on $S^2$ which leaves the vacuum fixed.
[We call its elements iso-rotations to distinguish them
from rotations in physical space].
Finally $P$ is a combined reflection in both space and the target space
$S^2$.

We are interested in stationary points of $\deg[\vec{n}]\ne 0$,
and the maximal subgroups of $G$ under which such fields can be invariant are
labelled by a nonzero integer $n$ and consist of spatial rotations by some
angle $\alpha\in [0,2\pi]$ and simultaneous iso-rotation by $-n\alpha$.
Fields invariant under such a group are of the form
\be
n_1=\sin f(\tilde{r})\,\cos (n\phi),\,\,\,\,\,\,\,\,\,
n_2=\sin f(\tilde{r})\,\sin (n\phi),\,\,\,\,\,\,\,\,\,
n_3=\cos f(\tilde{r})
\label{hed}
\ee
where $(\tilde{r},\phi)$ are polar coordinates and $f(\tilde{r})$
is the profile function.
Such fields are the analogues and generalizations of the hedgehog fields in 
the Skyrme model.
This parametrisation which involves azimuthal symmetry of the
fields assumes that all the skyrmions sit on top of each other while forming
the multiskyrmion configuration.

It is easy to show that the degree of the field (\ref{hed}) is
\be
\deg\left[\vec{n}\right]=n
\ee
ie equal to the azimuthal winding number $n$.

The corresponding static energy functional connected
with the Lagrangian (\ref{Lag}) for the OBM
and NBM is equal to
\begin{eqnarray}
E_{cl}(n)_{_{\tiny\mbox{OBM}}} & =& {g^2\over 2}\int r \,dr
\left(f'^2 + {n^2\sin^2 f\over r^2} +
a \biggl[{n^2f'^2 \sin f^2\over r^2}+2 \left(1-\cos f\right)\biggr]\right),
\label{en}\\
E_{cl}(n)_{_{\tiny\mbox{NBM}}}& =& {g^2\over 2}\int r \,dr
\left(f'^2 + {n^2\sin^2 f\over r^2} +
a \biggl[{n^2f'^2 \sin f^2\over r^2}+ \left(1-\cos\sp2 f\right)\biggr]\right)
\label{en1}
\end{eqnarray}
respectively.
In (\ref{en}) and (\ref{en1}) the length $(\sqrt{ge\mu})^{-1}$ has
 been absorbed so that the scale size of the localized structures
is a function of the dimensionless
spatial coordinate $r=\sqrt{ge\mu} \,\tilde{r}$ while the dimensionless 
parameter $a=\mu/ge$ becomes the only nontrivial parameter of the model.
Finiteness of the energy functional requires that the profile
function has to satisfy the following boundary conditions:
$f(0)=\pi$ and $f(\infty)=0$.

By setting $\phi=\cos f$ into (\ref{en}) the energy functional becomes
\be
E_{cl}(n)_{_{\tiny\mbox{OBM}}} = {g^2\over 2}\int r\,dr
\left({\phi'^2\over 1-\phi^2} +
{n^2\left(1-\phi^2\right)\over r^2}+
a \left[{n^2\phi'^2\over r^2}+2\left(1-\phi\right)\right]\right)
\ee
and a similar expression for $E_{cl}(n)_{_{\tiny\mbox{NBM}}}$.
Parametrizing the field $\phi$, using the ansatz introduced in
\cite{6} for the description of the three-dimensional skyrmions, as
\be
 \phi=\cos f = \frac{(r/r_n)^p-1}{(r/r_n)^p+1},
 \qquad  \, \qquad
\phi'={p\over 2r}(1-\phi^2)\label{ans}
\ee
leads after integration with respect to $r$ to the following analytic
energy expressions
\begin{eqnarray}
E_{cl}(n)_{_{\tiny\mbox{OBM}}}& =& \pi g^2 \left({4n^2\over
 p}+p+\frac{4a\pi}{p\,\sin(2\pi/p)}
\left[\frac{n^2(p^2-4)}{3r_n^2p}+r_n^2\right]\right),
\label{Ena}\\
E_{cl}(n)_{_{\tiny\mbox{NBM}}}& = &\pi g^2 \left({4n^2\over p}
+p+\frac{4a\pi}{p\,\sin(2\pi/p)}
\left[\frac{n^2(p^2-4)}{3r_n^2p}+\frac{2}{p}r_n^2\right]\right).
\label{Enaa}
\end{eqnarray}
Here $p$ and $r_n$ are parameters which still need to be determined
by minimizing the energy.
In fact, $r_n$ corresponds to the radius of the $n$th-soliton configuration.
{\large \it Remark:} The following Euler-type integrals have been used for the
derivation of (\ref{Ena}) and (\ref{Enaa}), see also \cite{6}\\
\begin{eqnarray}
\int_0^\infty \frac{2r\,dr}{1+(r/r_n)^p} &=&\frac{2\pi r_n^2}
{p\,\sin(2\pi/p)}, \,\,\,\,\, \, p>2\nonumber\\
\int_0^\infty \frac{dr\left(r/r_n\right)^p}{r\left[1+(r/r_n)^p\right]^2}
&=&{1\over p},\,\,\,\,\, \, p>0\nonumber\\
\int_0^\infty \frac{dr\left(r/r_n\right)^{2p}}
{r^3\left[1+(r/r_n)^p\right]^4} &=& \frac{\pi\left(p^2-4\right)}
{3r_n^2 \,p^4\,\sin(2\pi/p)},
\,\,\,\,  \, p>1 \nonumber\\
\int_0^\infty \frac{2r\,dr}{[1+(r/r_n)^p]\sp2} &=&\biggl(1-\frac{2}{p}\biggr)
\frac{2\pi r_n^2}
{p\,\sin(2\pi/p)}, \,\,\,\,\, \, p>1.
\end{eqnarray}
It can be easily proved that the minimization of the energies (\ref{Ena}) and
(\ref{Enaa}) implies that
\be
(r_n^{min})^2_{_{\tiny\mbox{OBM}}} =\frac{n}{\sqrt{3}}\sqrt{\frac{p^2-4}{p}},
\quad \quad \quad
(r_n^{min})^2_{_{\tiny\mbox{NBM}}}=n\sqrt{\frac{p\sp2-4}{6}}
\label{rmin}
\ee
ie $(r_n^{min})^2_{_{\tiny\mbox{OBM}}} =
\sqrt{{p\over 2}}\,(r_n^{min})^2_{_{\tiny\mbox{NBM}}}$
and so the minimum value of the energies is equal to
\begin{eqnarray}
E_{cl}(n)_{_{\tiny\mbox{OBM}}} &=&4\pi g^2\left[{n^2\over p}
+{p\over 4}+\frac{2an\pi} {\sqrt{3}p\sin(2\pi/p)}
\frac{\sqrt{p^2-4}}{\sqrt{p}}\right],
\label{apo}\\
E_{cl}(n)_{_{\tiny\mbox{NBM}}} &=&4\pi g^2\left[{n^2\over p}
+{p\over 4}+\frac{2\sqrt{2}an\pi}
{\sqrt{3}\sin(2\pi/p)}\frac{\sqrt{p^2-4}}{p\sp2}\right].
\label{apoa}
\end{eqnarray}
It is obvious that the energy contributions of the Skyrme and
the potential term are equal due to  (\ref{rmin}), which is in
agreement with the result obtained from Derrick's theorem.
Equations (\ref{apo},\ref{apoa}) provide an upper bound for the energies
of baby-skyrmions for any value of $p$. To get lower bound, we should
minimize the right-hand sides of (\ref{apo},\ref{apoa}) with respect to
the parameter $p$.
In what follows we investigate various cases which correspond to different
values of the only nontrivial parameter of the model, $a$.

First consider the case where $a \ll 1$, ie for very small values of
the model parameter.
Observe that for $a=0$ the ansatz (\ref{ans}) is a solution
of the model for $p=2n$, which implies that $p \rightarrow 2n$ as
 $a \rightarrow 0$.
In this case, due to (\ref{rmin}), the radius of the multiskyrmion
configuration increases with $n$:
$(r_n^{min})^2_{\tiny\mbox{OBM}} \sim n^{3/2}$ and
$(r_{n}^{min})^2_{\tiny\mbox{NBM}} \sim n^2$.
Moreover, the configuration consists of a ring of thickness:
$\delta\simeq 4r_n/p$ thus
$\delta_{\tiny\mbox{OBM}} \sim 2n^{-1/4}$ and
$\delta_{\tiny\mbox{NBM}} \sim \mbox{const}$.
{\large \it Remark}: The ring thickness is determined as the difference of the
  values of $\phi$ inside (which is equal to -1) and outside
(which is equal to +1)  of the ring (ie $d \phi =2$) divided by its
derivative at $r=r_n$ where, due to (\ref{ans}), $\phi(r_n)=0$
and so  $\phi'(r_n)=p/2r_n$.

This kind of magnetic solitons have been observed in \cite{11,12}
as solutions of the Landau-Lifshitz equations defining the dynamics of 
ferromagnets. [Note that the static solutions of the baby Skyrme model and the
 Landau-Lifshitz equations are related.]
In general, $\phi$ given by (\ref{ans}) for  $p=2n$ is a
low-energy approximation of multiskyrmion configurations (for $n>1$),
since for $n=1$ the corresponding energies given by (\ref{apo}) and (\ref{apoa}) are infinite.
Indeed, it is a matter of simple algebra to show that
\begin{eqnarray}
E_{cl}(n=2)_{_{\tiny\mbox{OBM}}}= 4\pi g^2 \left(2+a\pi\right),
 &\quad \quad &
 E_{cl}(n=2)_{_{\tiny\mbox{NBM}}}=4\pi g^2
\left(2+\frac{a\pi}{\sqrt{2}}\right),\nonumber\\
E_{cl}(n=3)_{_{\tiny\mbox{OBM}}}= 4\pi g^2
 \left(3+a\pi {8\over 3\sqrt{3}}\right), &&
E_{cl}(n=3)_{_{\tiny\mbox{NBM}}}=4\pi g^2
\left(3+a\pi {8\over 9}\right),\nonumber\\
E_{cl}(n=4)_{_{\tiny\mbox{OBM}}}=4\pi g^2
\left(4+a\pi \sqrt{5}\right),&&
E_{cl}(n=4)_{_{\tiny\mbox{NBM}}}=4\pi g^2
\left(4+a\pi {\sqrt{5}\over 2} \right).
\label{duo}
\end{eqnarray}
For large $n$, the energies take the asymptotic values
\begin{eqnarray}
E_{cl}(n)_{_{\tiny\mbox{OBM}}}&=& 4\pi ng^2 \biggl(1+
\sqrt{{2n \over 3}}a\biggr), \nonumber \\
E_{cl}(n)_{_{\tiny\mbox{NBM}}}&=&4\pi ng^2 \biggl(1+
\sqrt{{2 \over 3}}a\biggr).
\end{eqnarray}
Note that the energy of the OBM per unit
 skyrmion number increases as $n$ increases, while the energy of the NBM per
 skyrmion decreases with increasing $n$ to become constant for $n$ large.
In fact, the energies given by (\ref{duo}) are the upper bounds
of the multiskyrmion energies since the exact profile function corresponding
to the minimum of the energy differs from that given by (\ref{ans}).

\section{Perturbation Theory for the  Model Parameter}

In this section energy corrections up to  second or higher order
with respect to the model parameter $a$ have been obtained.
The corresponding energies for OMB and NBM  can be written as
\be
E_{cl}(n)=4\pi g^2[f(p)+a\,h(p)]
\label{egen}
\ee
where  $f(p)$ and $h(p)$ can be evaluated
 from (\ref{apo}) and (\ref{apoa}), respectively.
By letting $p=2n+\epsilon$ and expanding the  energies (\ref{apo})
and (\ref{apoa})  up to second  order in $\epsilon$  we
get $f(p)=n+\epsilon^2/(8n)$ and $h(p)=h_0+\epsilon h_1$ where
$h_1= (2n)^{-1} \beta h_0$.
In fact, the corresponding functions for the OBM and the
NBM are given by
\begin{eqnarray}
\frac{h_{0_{\tiny \mbox{OBM}}}}{n} =
\sqrt{{2n\over 3}}\frac{\pi}{n\,\sin(\pi/n)}\sqrt{1-1/n^2},
&\qquad&
\beta_{\tiny \mbox{OBM}} = \frac{\pi}{n}\cot(\pi/n)-{1\over 2}+{1\over n^2-1},
\label{hbold}\\
\frac{h_{0_{\tiny \mbox{NBM}}}}{n} =\sqrt{{2\over 3}}\frac{\pi}
{n\,\sin(\pi/n)}\sqrt{1-1/n^2},
&&
\beta_{\tiny \mbox{NBM}} = \frac{\pi}{n}\cot(\pi/n)\,-\,1\, +{1\over n^2-1}.
\label{hbnew}
\end{eqnarray}

Minimization of  (\ref{egen}) with respect to $\epsilon$ implies that
$\epsilon^{min}=-4anh_1=-2a\beta h_0$.
At large values of $n$, the parameteres $\epsilon$ and $p=2n+\epsilon$
take the values
\begin{eqnarray}
\epsilon(n)_{_{\tiny \mbox{OBM}}} \simeq -an \sqrt{{\frac{2n}{3}}},
&\qquad& \epsilon(n)_{_{\tiny \mbox{NBM}}}\simeq
2a\sqrt{{2\over 3}}\frac{\pi^2/3-1}{n}
\label{eps}\\
p(n)_{_{\tiny \mbox{OBM}}}\simeq 2n-an\sqrt{{2n\over 3}},
&\qquad&
p(n)_{_{\tiny \mbox{NBM}}} \simeq 2n\,+2a\sqrt{{2\over 3}}\frac{\pi^2/3
-1}{n}.
\label{pow}
\end{eqnarray}
For any $a$, as $n$ increases, the effective power
$p(n)_{_{\tiny \mbox{OBM}}}$ becomes negative and the approach based
on the  assumption that
$\epsilon_{_{\tiny \mbox{OBM}}}$ is small  is not self-consistent (also, see next section).
On the contrary, for NBM,
 $p(n)_{_{\tiny \mbox{NBM}}} \simeq 2n$ as $n$ increases
 which implies that our consideration  is self-consistent in this case.
In terms of (\ref{egen})-(\ref{hbnew}), the energy per skyrmion of the $n$-th 
skyrmion configuration takes the value
\be
\frac{E_{cl}(n)}{4\pi g^2n}=1+a{h_0\over n}-a^2\frac{h_0^2\beta^2}{2n^2},
\label{eunit}
\ee
which gives us
\begin{eqnarray}
\frac{E_{cl}(2)_{\tiny \mbox{OBM}}}{4\pi g^2 2} =1\,+1.5708 \,a\,-0.034\,a^2,
&\quad \quad \quad&
\frac{E_{cl}(2)_{\tiny \mbox{NBM}}}{4\pi g^2 2} =1\,+1.1107\,a\,-0.2741\,a^2,
\nonumber \\
\frac{E_{cl}(3)_{\tiny \mbox{OBM}}}{4\pi g^2 3} =1\,+1.6120 \,a\,-0.068\,a^2,&&
\frac{E_{cl}(3)_{\tiny\mbox{NBM}}}{4\pi g^2 3} =1\,+0.9308\,a\,-0.0317\,a^2,
\nonumber \\
\frac{E_{cl}(4)_{\tiny \mbox{OBM}}}{4\pi g^2 4} =1\,+1.7562 \,a\,-0.191\,a^2,&&
\frac{E_{cl}(4)_{\tiny\mbox{NBM}}}{4\pi g^2 4} =1\,+0.8781\,a\,-0.0084\,a^2
\nonumber \\
\frac{E_{cl}(5)_{\tiny \mbox{OBM}}}{4\pi g^2 5} =1\,+1.9122 \,a\,-0.302\,a^2,&&
\frac{E_{cl}(5)_{\tiny\mbox{NBM}}}{4\pi g^2 5} =1\,+0.8552\,a\,-0.0032\,a^2,
\nonumber \\
\frac{E_{cl}(6)_{\tiny \mbox{OBM}}}{4\pi g^2 6} =1\,+2.0649 \,a\,-0.404\,a^2,&&
\frac{E_{cl}(6)_{\tiny\mbox{NBM}}}{4\pi g^2 6} =1\,+0.8430\,a\,-0.0015\,a^2.
\label{enumold}
\end{eqnarray}

For large $n$, the energies (\ref{eunit}) take the asymptotic values
\begin{eqnarray}
\frac{E_{cl}(n)_{_{\tiny \mbox{OBM}}}}{4\pi g^2n}&=&\left(1\,
+a\sqrt{{2n\over 3}}\,-\,a^2{n\over 12}\right),\nonumber\\
\frac{E_{cl}(n)_{_{\tiny \mbox{NBM}}}}{4\pi g^2n}&=&\left(1\,
+a\sqrt{{2\over 3}}\,-\,a^2\frac{(\pi^2/3-1)^2}{3n^4}\right).
\end{eqnarray}
Note that the energies of the two models behave differently
when we consider terms of the second order in the model parameter,
ie terms $ \sim a^2$.
Indeed, for the OBM the contribution to the energy
is linearly proportional to the skyrmion number $n$, while for the
NBM the contribution decreases rapidly as the skyrmion number increases.
This implies that the linear approximation in $a$ is accurate for the
NBM since the quadratic term becomes negligible for large $n$.
Numerical results obtained for different values of $a$
for the OBM and NBM are presented in {\it Table 1} and {\it Table 2}, 
respectively.

As we have pointed out earlier, our method cannot describe the
one-skyrmion configuration since the corresponding energies
become infinite.
However, by setting $p=2+\varepsilon$ to (\ref{apo}) and (\ref{apoa})
and expanding all terms up to third order in $\varepsilon \ll 1 $ we obtain
\be
E_{cl}(n=1) = 4\pi g^2\left( 1 +{\varepsilon^2\over 8}-
{\varepsilon^3\over 16}
 +2a\sqrt{{2\over 3\varepsilon}}(1\,-\gamma \varepsilon)\right).
\label{n1exp}
\ee
where $\gamma$ has different value for each model, ie
\be
\gamma_{\tiny\mbox{OBM}} \,=\, {1\over 8},
\qquad \qquad
\gamma_{\tiny\mbox{NBM}} \,=\, {3 \over 8}.
\label{gamma}
\ee
Note, that when only terms up to second order in $\varepsilon$
have been considered, the corresponding energy (\ref{n1exp}) simplified to:
$E_{cl}= 4\pi g^2\left( 1 +{\varepsilon^2\over 8}
+2a\sqrt{{2\over 3\varepsilon}}\right)$, and the minimum occurs at:
$\varepsilon_{_{1}} = 2(a/\sqrt{3})^{2/5}$.
Finally, the minimum of (\ref{n1exp}) occurs at
\be
\varepsilon^{min} =\, 2\left({a\over \sqrt{3}}\right)^{2/5}\left[1\,+\,
{4\over 5}\left({a\over \sqrt{3}}\right)^{2/5}\left(\gamma + {3\over 4}\right)
\right]
\label{eps1}
\ee
and corresponds to a shift of  $\varepsilon_{_{ 1}}$ since higher order
corrections of $\varepsilon$ have been considered in (\ref{n1exp}).
The energy of the one-skyrmion configuration is
\begin{eqnarray}
\frac{E_{cl}(n=1)}{4\pi g^2}& = & \left\{1 \,+{5\over 2}\left({a\over \sqrt{3}}
\right)^{4/5}\left[1\,-\,{1\over 5}\left({a\over \sqrt{3}}\right)^{2/5}
(8\gamma +1)\right]\right\} \nonumber\\
 &\simeq &\left[1+1.611\, a^{4/5}
\left(1-\,0.1605 \,a^{2/5}\left(8\gamma+1\right)\right)\right].
\label{ena1}
\end{eqnarray}
Equation (\ref{ena1}) implies that, for a single skyrmion, the energy
expansion in $a$ is proportional to a power of $a$ instead of being
linearly proportional to $a$ (which is the case for the multiskyrmion 
configurations with $n \geq 2$), while its convergence is worse than for 
multiskyrmions,  especially for the NBM.
In fact, when $a=0.4213$ the first two terms in (\ref{ena1}) are equal to
$1.807$, while the next order term lowers this value down to $1.44$,
which compared with the exact value $1.564$ obtained from numerical 
simulations gives an error of 7\%.
Note, that our one-skyrmion parametrization gives the same energy for both
models, when expansions only up to the lowest order in $a$ have been 
considered: the difference appears only in the term $\sim a\gamma 
\sqrt{\varepsilon} $ in (\ref{n1exp}).

\begin{center}
\begin{tabular}{|l|l|l|l|l|l|l|l|}
\hline
$a$/$n$ &$n=1$ &$n=2 $&$n=3  $&$n=4 $ &$n=5 $&$n=6$&$n=8$ \\
\hline
$a=0.001$&$1.0063$&$1.00157$&$1.0016$&$1.0017$&$1.0019$&$1.0021$&$1.0023$\\
\hline
$a=0.01$&$1.0384$&$1.0157$&$1.0161$&$1.0176$&$1.0191$&$1.0206$&$1.0234$\\
\hline
$a=0.0316$&$1.0933$&$1.0496$&$1.0508$&$1.0553$&$1.0601$&$1.0649$&$1.0737$\\
\hline
$a=0.1$&$1.2227$&$1.1567$&$1.1605$&$1.1737$&$1.1882$&$1.2025$&$1.2291$ \\
\hline
$a=0.316$    &$1.5113$&$1.4930$&$1.5026$&$1.5358$&$1.5638$&
$1.6126$&$1.6835$ \\
\hline
\hline
$a_{\mbox{\tiny hed}}=0.316 \,(\mbox{num})$
&$1.5647$&$1.4681$&$1.4901$&$1.5284$&$1.5692$&$1.6092$&$1.6832$ \\
\hline
$a=0.316 \,(\mbox{num})$ &$1.564$&$1.468$&$1.460$&$1.450$&$1.456$&$1.449$&$-$\\
\hline
\end{tabular}
\end{center}
{\tenrm
\baselineskip=11pt

{\it Table 1:} Energy per unit skyrmion number (in $4\pi g^2$) for different 
values of the parameter $a$ for the OBM case where corrections of second 
order in $a$ have been taken into account.
The last two lines contain the exact results obtained from the numerical
simulations of multiskyrmions ($n \geq 2$) with ring-like shapes
and ($n\geq 3$) with shapes other than ring-like \cite{10}, respectively.
For the first case, we have solved numerically the equations using 
the hedgehog ansatz (\ref{hed}).}\\

\begin{center}
\begin{tabular}{|l|l|l|l|l|l|l|l|l|l|}
\hline
$a$/$n$ &$n=1$   &$n=2 $  &$n=3  $ &$n=4 $  &$n=5 $  &$n=6$ &$n=8$&$n=12$
  &$n=16$ \\
\hline
$a=0.01$&$1.0363$&$1.0111$&$1.0093$&$1.0088$&$1.0085$&$1.0084$&$1.0083$&
$1.0082$&$1.0082$\\
\hline
$a=0.0316$&$1.0851$&$1.0348$&$1.0294$&$1.0277$&$1.0270$&$1.0266$&$1.0262$&
$1.0260$&$1.0259$\\
\hline
$a=0.1$&$1.1887$&$1.1083$&$1.0928$&$1.0877$&$1.0855$&$1.0843$&$1.0831$&
$1.0823$&$1.0820$ \\
\hline
$a=0.316$ &$1.3814$&$1.3238$&$1.2912$&$1.2768$&$1.2699$&$1.2662$&$1.2626$&
$1.2602$&$1.2593$ \\
\hline
$a=0.4213$ & $1.44$ &$1.4193$&$1.3865$&$1.3684$&$1.3597$&$1.3549$&$1.3501$&
$1.3467$&$1.3455$ \\
\hline
\hline
$a=0.4213 \,(\mbox{num})$ &$1.564$&$1.405$&$1.371$&$1.358$&$1.352$&$1.349$&
$1.3447$&$1.3407$&$1.3385$ \\
\hline
\end{tabular}
\end{center}
{\tenrm
\baselineskip=11pt

{\it Table 2:} Energy per unit skyrmion number for different values of the
parameter $a$ for the NBM.
The last line contains the exact results determined by the numerical
simulations \cite{10} of multiskyrmions with ring-like shapes
when $a=0.4213$, for $n\leq 6$ coinciding with ours.}\\

By looking at the results presented in {\it Table 1} and
{\it Table 2} it is clear that
our approximate method gives the energy values which are quite close to the
exact values obtained by numerical simulations, especially for the NBM.
In particular, the difference between the exact and approximate energy
for $a=0.4213$ is less than 0.5\% for $n \geq 6$. For smaller values of $a$
the agreement between analytical and numerical results is even better.
In evident agreement with (\ref{poten}), the energies of the NBM skyrmions 
given in {\it Table 2} are smaller than those of the OBM skyrmions 
(see {\it Table 1}) at the same values of the model parameters.

Note that, for the OBM (when  $a$ is  small)
 the energy per skyrmion of a multiskyrmion
configuration with $n\geq 2$ is smaller compared to the single skyrmion energy
 and therefore, these configurations are bound states, stable with
respect to the decay into $n$ individual skyrmions.
On the contrary,  the ring-like OBM multiskyrmions with even $n$ 
(where $n \geq 4$)
are unstable with respect to the decay into two-skyrmion configurations,
while configurations with odd $n$ (where $n \geq 5$)
are unstable with respect to the breakup into a two- and  a three-skyrmion
configurations.
In addition, {\it Table 1} and (\ref{ena1}) show that for any $n$ 
(where $n \neq 1$) there is an upper limit for the model 
parameter: $a\le a_{cr}(n)$ above which the $n$-th ring-like skyrmion 
configuration can decay into $n$ individual skyrmions.

Let us consider the case of $n=3$ in more detail. As it can be observed from 
the energies (\ref{duo}) and (\ref{ena1})
when $a \leq 0.77$  the ring-like three-skyrmion configuration is
stable with respect to the decay into a single and a two-skyrmion configuartion
since
\be
E_1+E_2 -E_3 \simeq 1.611 \,a^{4/5} - a\pi \left(\frac{8}{3\sqrt{3}} -1\right),
\ee
and this difference  becomes positive when and only when
\begin{eqnarray}
a &\leq& \left(\frac{3\sqrt{3}\,1.611 }{\pi\left(8-3\sqrt{3}\right)}
\right)^5\nonumber\\
&\simeq &\, 0.77.
\end{eqnarray}
Corrections to the energy for the skyrmion
configurations with $n=1,2,3$ of the higher order in $a$, lead to smaller 
critical values $a_{cr}(n)$.

Since our fields with axial symmetry (\ref{hed}) and (\ref{ans}) correspond to
ring-like solutions of the Euler-Lagrange equations \cite{1} for $a=0$, they
have to be solutions of the corresponding equations also as $a \rightarrow 0$,
ie when $a$ takes values in a small region close to zero.
[In fact, this region becomes more narrow as $n$ increases 
as in this limit the expansion in $a$ becomes less convergent].
On the other hand, the lattice-like configurations (tripole
for $n=3$, quadrupole for  $n=4$,  etc.) are solutions of
the equations when $a \geq a_{cr}(n)$ for given $n$ \cite{5,10,13}.
However, the transition from the ring-like configuration to any other 
minimal energy configuration is a  phenomenon which has not been studied 
in much detail yet and deserves further investigation.

Finally, it should be stressed that, in contrast to the linear approximation,
the quadratic approximation given by (\ref{enumold}) does not provide
an upper bound for the energy.

\section{Away from the Nonlinear $O(3)$ $\sigma$-Model }

In the general case, for arbitrary values of the parameter $a$ and the
skyrmion number $n$,
soliton solutions can be obtained by minimizing numerically the energy
(\ref{apo}) and (\ref{apoa}) with respect to the variable $p$.
This way an upper bound is obtained for the corresponding energies
since the profile function is given by (\ref{ans}).

For large $a$ at fixed $n$ (or for large $n$ at fixed $a$),
 the expansion (\ref{hbold}) is not self-consistent for the OBM.
However, some analytical results can also be obtained in this case
since (\ref{apo}) for  large $a$ can be  approximated by
\be
E_{cl}(n)_{_{\tiny\mbox{OBM}}} \simeq 4\pi g^2 \,\frac{2an\pi}
{\sqrt{3} p\,\sin(2\pi/p)}\frac{\sqrt{p^2-4}}{\sqrt{p}}.
\label{away}
\ee
Expansion, up to second order terms of (\ref{away}) with respect to $p$,
gives
\be
E_{cl}(n)_{_{\tiny\mbox{OBM}}}\simeq 4\pi g^2
\frac{an\sqrt{p}}{\sqrt{3}}\left(1\,+{c_2\over p^2}\,
\right)
\label{awayap}
\ee
where $c_2=2(\pi^2/3-1)$; while its minimization implies
that $p_{min}\simeq \sqrt{3c_2}=3.71$ and so the corresponding energy is
\begin{eqnarray}
\frac{E_{cl}(n)_{_{\tiny\mbox{OBM}}}}{4\pi g^2}&\simeq &
{4\over 3}an \left({c_2\over 3}\right)^{1/4}\nonumber\\
&=&\, 1.48\, an
\end{eqnarray}
Note that, by contrast with the results obtained near the
nonlinear O(3) $\sigma$-model, for large $a$ the parameter $p$ is independent
of the skyrmion number $n$.
Also, for  $a\gg n$ the skyrmion radius is proportional to
the square root of the skyrmion number: $r_n  \sim n^{1/2}$; while the skyrmion
thickness is given by: $\delta\sim r_n/p \sim n^{1/2}$, and so the ring-like
structure of the configuration is not very pronounced.

Direct numerical minimization of (\ref{away}) with respect to $p$ gives 
$p_{min}=4.5$ and the corresponding value of the energy is
\be
\frac{E_{cl}(n)_{_{\tiny\mbox{OBM}}}}{4\pi g^2}=1.55\,an.
\ee
The energy, which has been obtained by solving numerically
the Euler-Lagrange equation \cite{8}, is:  $E_{cl}(n)/4\pi g^2 = 1.333\, an$.
The profile function corresponding to this solution is given by: 
$\cos f ={r^2\over 8n^2}(r_n^2-r^2)+2{r^2\over r_n^2} -1$ for $r\leq r_n$ and 
$f=0$ for $r> r_n$.
This solution is quite different from our parametrization (\ref{ans}) and thus,
the $16\%$
difference between the exact and the approximate solution is understandable.

To conclude, we recall that for NBM
the parametrization (\ref{ans}) works well for arbitrary large $n$,
and its accuracy increases  with increasing $n$, as illustrated in 
{\it Table 2}.

\section{Properties of the Skyrmions: Mean Square Radii, Energy Density, Moment
of Inertia}

Many properties of  multiskyrmions can be determined using the ansatz (\ref{ans}).
For example, the mean square radius of the $n$-th multiskyrmion configuration
takes the simple form
\begin{eqnarray}
<r^2>_n &=& {1\over 2}\int dr\, r^2 \phi' \nonumber\\
&=&\frac{2\pi r_n^2}{p \sin(2\pi/p)}
\label{rad}
\end{eqnarray}
where $r_n$ is given by (\ref{rmin}) for the OBM and NBM.
For small $a$, it was shown in section 2 that $p=2n$,
which implies that the mean square radius becomes
\be
<r^2>_{n_{_{\tiny\mbox{OBM}}}}\simeq \frac{\pi\sqrt{2(n^2-1)}}
{\sin(\pi/n)\sqrt{3n}}, \qquad \qquad
<r^2>_{n_{_{\tiny\mbox{NBM}}}}\simeq \frac{\pi
\sqrt{2(n^2-1)}}{n\sin(\pi/n)\sqrt{3}}
\ee
which is equal to $\pi,\;8\pi/3\sqrt{3},\; \pi\sqrt{5},\dots$ and
$\sqrt{2}\pi,\;8\pi/3,\; 2\pi\sqrt{5},\dots$ for $n=2,3,4,\dots$,
respectively.

For the NBM, even for a large enough value of the parameter $a$, the analytical formula
(\ref{rmin}) with the power $p$ taken from  (\ref{pow}) gives the values
of $<r^2>_{n_{\tiny \mbox{NBM}}}$ in a remarkably good agreement with those obtained in numerical 
calculations.
E.g., for $n=3$ the analytical result is $\sqrt{<r^2>_3} = 2.987$, in natural 
units of the model $1/(ge\mu)$, to be compared with $2.872$ obtained numerically. 
This agreement improves with increasing $n$, and for $n=12$ we have
$\sqrt{<r^2>_{12}} \sim 10.92$ to be compared with $10.85$ determined numerically. A similar agreement
between analytical and numerical results takes place for the mean square 
radius of the energy distribution of multiskyrmions (the 3D-case has been 
considered in detail in \cite{6}).

Note that the one-skyrmion configuration is (still) a singular case since
 (\ref{rad}) is not defined for $n=1$.
However, as we have shown earlier, by expressing $p=2+\varepsilon$
and expanding (\ref{rmin}) in  $\varepsilon$ we get $r_{n=1}^2 =
\sqrt{2\varepsilon/3}$  which leads to
\be
<r^2>_1 = 2\sqrt{\frac{2}{3\varepsilon^{min}}}
\label{r1}
\ee
for $\varepsilon^{min}$ given by (\ref{eps1}).
So, our approximate method shows that as the model parameter tends to zero
the mean square radius of the one skyrmion field tends to infinity since
$<r^2>_1 \sim a^{-1/5}$; while since $<r^2>_{_{\tiny\mbox{NBM}}}(n)=
\sqrt{n}<r^2>_{_{\tiny\mbox{OBM}}}(n)$ in this case the mean square radius
is given by (\ref{r1}) for both models.

The average energy density per unit surface element is defined as
\be
\rho_{_{\tiny E}} = \frac{E_{cl}(n)}{2\pi r_n\delta}.
\label{den}
\ee
with $\delta \simeq 2r_n/n$, see discussion after (\ref{apoa}). For the NBM, 
when $n$ is large, (\ref{den}) takes the constant value
\be
\rho_{_{E_{\tiny \mbox{NBM}}}}\simeq e\mu\, g^3\left(\sqrt{{3\over 2}} 
+a\right)
\label{denn}
\ee
ie is independent of $n$. So, (\ref{denn}) represents the fundamental 
property of this kind of multiskyrmions.
On the contrary, for the OBM when ring-like configurations (which do not 
correspond to the minimum of the energy \cite{5,10}) are considered, the
energy density increases with $n$ like $\sim \sqrt{n}$ at small values of $a$.

Another quantity of physical significance determining the quantum corrections to
the energy of skyrmions is the moment of inertia which has been considered
for two-dimensional models in \cite{13}.
In order to obtain the energy quantum correction of the soliton,
due to its rotation around the axis perpendicular
to the plane in which the soliton is located, we have to take the $t$-dependent ansatz
of the form
\be
n_1=\sin f(\tilde{r})\,\cos [n(\phi-\omega t)],\,\,\,\,\,\,\,\,\,
n_2=\sin f(\tilde{r})\,\sin [n(\phi-\omega t)],\,\,\,\,\,\,\,\,\,
n_3=\cos f(\tilde{r}).
\label{hedom}
\ee
Then the $\omega$-dependence of the energy is given by the simple formula:
\be
E^{rot} = {\Theta_J \over 2} \omega^2
\ee
where  $\Theta_J$, the so-called moment of inertia, is given by \cite{13}
\be
\Theta_J(n) = g^2n^2\int d^2r\, \sin^2f \, \left(1+a f'^2\right).
\label{tj}
\ee
Using (\ref{ans}) and the relations
\begin{eqnarray}
{1\over 4}\int (1\,-\,\phi^2)r\,dr &=&
\int_{0}^{\infty} \frac{\left(r/r_n\right)^p r\,dr}
{\left[1+\left(r/r_n\right)^p\right]^2}\, \nonumber\\
 &=&
\frac{2\pi\,r_n^2}{p^2\sin(2\pi/p)},  \,\,\,\,\, \, p>2\nonumber\\
{1 \over 16}\int (1\,-\,\phi^2)^2\; {dr\over r} &=&
\int_{0}^{\infty} \frac{\left(r/r_n\right)^{2p} \,dr}{\left[1
+\left(r/r_n\right)^p\right]^4r}\, \nonumber\\
 &=& {1\over 6p}, \,\,\,\,\, \, p>0
\end{eqnarray}
we find that at large values of $n$ the moment of inertia simplifies to
\be
\Theta_J(n)\,\simeq 4\pi g^2nr_n^2\left({2n\over p} +{anp\over 3r_n^2}\right)
\label{tjf}
\ee
 which holds for any multiskyrmion configuration described by ansatz
(\ref{ans}), for both models.
For small values of $a$, letting $p=2n$ and taking $r_n^2$ given by 
(\ref{rmin}) we find that
\begin{eqnarray}
\Theta_J(n)_{_{\tiny \mbox{OBM}}}&\simeq & 4\pi g^2 n\,r_n^2
\left(1 +\,a\sqrt{{2n\over 3}}\right)
\nonumber\\
\Theta_J(n)_{_{\tiny \mbox{NBM}}} &\simeq & 4\pi g^2 n\,r_n^2 \left( 1
+ a\sqrt{{2\over 3}}\right),
\label{tjn}
\end{eqnarray}
which implies that the moment of inertia, for large $n$, is
\be
\Theta_J(n) \simeq E_{cl}(n)\, r_n^2,
\ee
in agreement with simple quasi-classical arguments for the thin massive ring.
Similar quasi-classical formulae have been obtained for the three-dimensional
skyrmions (see, for example, Ref. \cite{6,14}) where the moment of inertia
was shown to be given by $\Theta_J = 2 M_B r_B^2/3$ for large baryon numbers; 
this expression is valid for a classical spherical bubble with the mass 
concentrated in its shell.

\section{Conclusions}

In this paper we have presented an analytical approach for deriving
approximate expressions to skyrmion solutions of the two-dimensional $O(3)$
$\sigma$-model.
These approximations are very accurate for small values of the parameter $a$ 
which determines the weight of the Skyrme and the potential term in the 
Lagrangian. For other values
of the model parameter we have performed some
numerical calculations and then combined them with further analytical 
work to investigate the binding and other properties of multiskyrmion states.

Two models have been studied: the ``{\it old baby Skyrme model}" and
the ``{\it new baby Skyrme model}" which differ from each other in the form of 
the  potentials (\ref{poten}).
For both models the $a$ dependence of the energy of a single skyrmion differs
from the cases where topological number $n\geq 2$.
For the OBM, when $a$ is small,
the $n=3$ skyrmion configuration is stable with respect to the decay into a
single skyrmion and a two-skyrmion configuration while the ring-like 
multiskyrmion configurations with $n \geq 4$ are unstable with respect to the 
breakup into two- and three-skyrmion configurations.

For the NBM, on the other hand, the hedgehog multiskyrmion configurations 
considered in \cite{10} and here describe bound states
 since the energy per skyrmion decreases as the skyrmion number increases.
We note that the results obtained for the NBM are similar to
 the ones obtained for the three-dimensional model studied in \cite{6}.
In both cases the energy per skyrmion decreases as the skyrmion number
increases. The three-dimensional skyrmions obtained using the rational
map ansatz \cite{15}, for large $n$, have the form of a
bubble with energy and baryon number concentrated in the shell.
The thickness and the energy density of the shell (which is analogues to the
thickness of the ring in the two-dimensional case) do not depend
on the skyrmion number \cite{6}.
Similarly, in this paper we have shown that the two-dimensional baby skyrmions
of the NBM, for large $n$, correspond to
ring-like configurations with constant thickness and constant energy density
per unit surface of the ring. The building material for these objects is a band
of matter with constant thickness and average energy density per unit surface.
Thus the baby skyrmions can be obtained as dimensional reductions
of the three-dimensional skyrmions at large $n$; while the three-dimensional
skyrmions can be derived from the two-dimensional baby skyrmions as
dimensional extensions.

In \cite{8} it has been concluded that the Casimir energy, or quantum loop
corrections, can
destroy the binding properties of the two-skyrmion bound states.
It would be worth to investigate the validity of this argument for
the two- and three-skyrmion bound states of the NBM.
Another interesting question is to determine to what extent the region of 
small enough $a$ is of importance from the point of view of physics.
For large $a$ the method overestimates the skyrmion masses for 
the OBM but is accurate for the NBM, especially for large $n$.

The existence of bound states of the three-dimensional skyrmions
has rich phenomenological consequences in elementary particles and nuclear
physics. It suggests possible existence of multibaryons with nontrivial flavor,
 strangeness, charm or beauty;  more details are given in \cite{14} and
references therein.
Similarly, the existence of bound states of two-dimensional baby skyrmions
 with universal properties in the NBM, which describes
anisotropic systems,  can also have some consequences for the condensed state
physics, which would be worth to investigate in detail. \\

VBK is indebted to G. Holzwarth for drawing his attention to
the paper \cite{8}; his work is supported by the Russian Foundation for
Basic Research, grant 01-02-16615.
TI thanks the Nuffield  Foundation for a newly appointed lecturer award.
\\

{\elevenbf\noindent References}
\vglue 0.1cm


\begin{thebibliography}{20}
\tenrm\baselineskip=11pt
\bibitem{1} A.A. Belavin and A.M. Polyakov, JETP Lett. 22, 245 (1975)
\bibitem{2} R.A. Leese, M. Peyrard and W.J. Zakrzewski, Nonlinearity 3,
387 (1990)
\bibitem{3} B.M.A.G. Piette and W.J. Zakrzewski, Nonlinearity 9, 897 (1996)
\bibitem{4} T.A. Ioannidou, Nonlinearity 10, 1357 (1997)
\bibitem{5} B. Piette, B. Schroers and W.J. Zakrzewski, Z. Physik, C65,
165 (1995)
\bibitem{6} V.B. Kopeliovich, JETP Lett. 73, 587 (2001);
hep-ph/0109229, J.Phys.G, 28, 103 (2002)
\bibitem{7} A.E. Kudryavtsev, B.M.A.G. Piette and W.J. Zakrzewski,
Nonlinearity 11, 783 (1998)
\bibitem{8} H. Walliser and G. Holzwarth, hep-ph/9905492
\bibitem{9} G.S. Adkins and C.R. Nappi, Nucl. Phys. B 233, 109 (1984)
\bibitem{10} T. Weidig, hep-th/9811238; Nonlinearity 12, 1489 (1999);
T.Weidig, hep-th/9911056
\bibitem{11} A.M. Kosevich, B.A. Ivanov and A.S. Kovalev, Phys. Rept. 194, 117
(1990)
\bibitem{12} B. Piette and W. Zakrzewski, Physica D119, 314 (1998)
\bibitem{13} B. Piette, B. Schroers, and W.J. Zakrzewski, Nucl.Phys. B439,
205 (1995)
\bibitem{14} V.B. Kopeliovich, JETP 93, 435 (2001)
\bibitem{15} C.J. Houghton, N.S. Manton and P.M. Sutcliffe, Nucl. Phys. B510,
507 (1998)
\end{thebibliography}
\end{document}